\documentclass{article}

\usepackage{arxiv}

\usepackage[utf8]{inputenc} % allow utf-8 input
\usepackage[T1]{fontenc}    % use 8-bit T1 fonts
\usepackage{hyperref}       % hyperlinks
\usepackage{url}            % simple URL typesetting
\usepackage{booktabs}       % professional-quality tables
\usepackage{amsfonts}       % blackboard math symbols
\usepackage{nicefrac}       % compact symbols for 1/2, etc.
\usepackage{microtype}      % microtypography
\usepackage{lipsum}		% Can be removed after putting your text content
\usepackage{graphicx}
\usepackage{natbib}
\usepackage{doi}

\title{Multi-Stage Threat Modelling and Security Monitoring in 5GCN}

\author{ 

\href{https://orcid.org/0000-0000-0000-0000}{\includegraphics[scale=0.06]{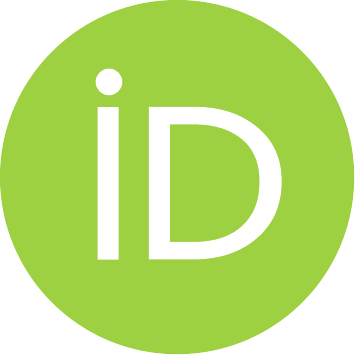}\hspace{1mm}Robert Pell}\\
	Department of Computer Science\\
	University of Surrey, UK\\ \\
	\texttt{r.pell@surrey.ac.uk} \\
	\And
	Sotiris Moschoyiannis\\
	Department of Computer Science\\
	University of Surrey, UK\\ \\
	\texttt{s.moschoyiannis@surrey.ac.uk} \\
	
	\And
	\href{https://orcid.org/0000-0001-7306-4062}{\includegraphics[scale=0.06]{orcid.pdf}\hspace{1mm}Emmanouil Panaousis} \\
	School of Computing and Mathematical Sciences\\
	University of Greenwich\\
	\texttt{e.panaousis@greenwich.ac.uk} \\
}

\renewcommand{\shorttitle}{\textit{arXiv} Template}

\hypersetup{
pdftitle={Multi-Stage Threat Modelling and Security Monitoring in 5GCN},
pdfsubject={q-bio.NC, q-bio.QM},
pdfauthor={Robert Pell, Emmanouil Panaousis, Sotiris Moschoyiannis},
pdfkeywords={5G Networks, Threat Modelling, Attack Graphs},
}

\begin{document}
\maketitle

\begin{abstract}
	The fifth generation of mobile networks (5G) promises a range of new capabilities including higher data rates and more connected users. To support the new capabilities and use cases the 5G Core Network (5GCN) will be dynamic and reconfigurable in nature to deal with demand. It is these improvements which also introduce issues for traditional security monitoring methods and techniques which need to adapt to the new network architecture. The increased data volumes and dynamic network architecture mean an approach is required to focus security monitoring resources where it is most needed and react to network changes in real time. When considering multi-stage threat scenarios a coordinated, centralised approach to security monitoring is required for the early detection of attacks which may affect different parts of the network. In this chapter we identify potential solutions for overcoming these challenges which begins by identifying the threats to the 5G networks to determine suitable security monitoring placement in the 5GCN.
\end{abstract}

% keywords can be removed
\keywords{5G Security \and threat modelling \and multi-stage attacks \and security monitoring}

\section{Introduction}
Security analysis of computer networks has become a vital tool to identify security risks, evaluate vulnerabilities and guide the decisions on protecting the network. In recent years networks have grown in complexity leading to the development of security models to conduct effective security analysis \cite{bodeau2018cyber}. A prime example of the evolution in network complexity is 5G mobile networks. 5G will incorporate a range of new technologies including Software Defined Networking (SDN), Network Function virtualisation (NFV), Multi-access Edge Computing (MEC) and transition to a cloud based service. All of these technologies provide the capability for 5G networks to meet the performance requirements and support a variety of new use cases but they also introduce an increased attack surface and a range of new threats.

\begin{figure}[h]
\includegraphics[width=\textwidth]{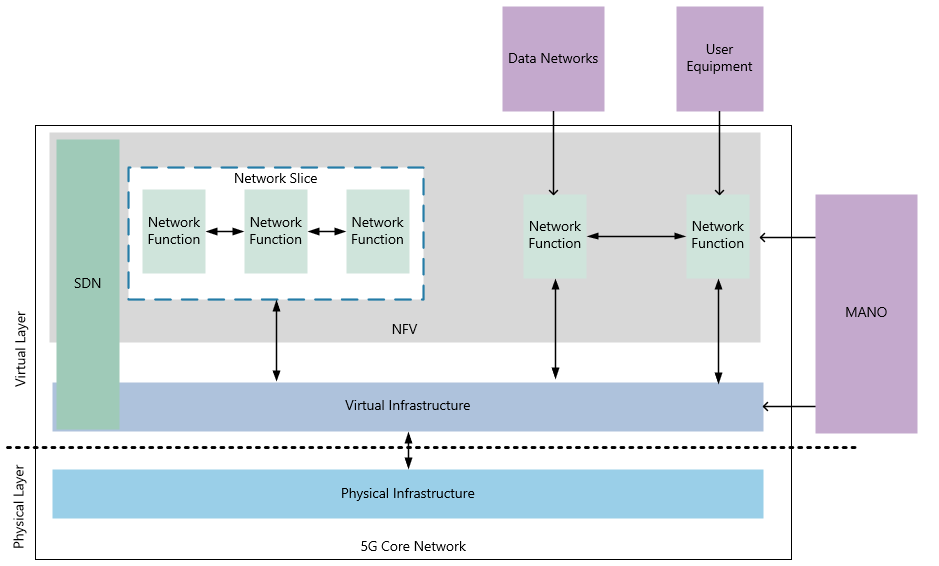}
\caption{Overview of the 5G Network architecture}
\centering
\label{fig:Architecure}
\end{figure}

The 5GCN architecture is still under development and is currently in the specification stage \cite{3GPP5GArchitecture}. Specific details and industry wide deployments of the 5GCN may not be realised for several years but the overarching architecture and the key components of the 5GCN are defined as shown in Fig.\ref{fig:Architecure}. It is the key building blocks demonstrated in this diagram which are the focus of various threat assessments to date. The building blocks of the 5GCN inherit many existing technologies and there are plenty of examples of virtualized, cloud based deployments existing across different industries. What makes the 5GCN a unique challenge from a security perspective is the radical transition from previous generation of mobile networks. The transition includes the introduction of Network Slicing (NS), the move away from centralised hardware and unique interfaces to different user equipment (UE), 3rd party applications and other mobile network service providers. Each of these interfaces are critical for providing the required functionality of telecommunication networks but they also introduce an attack surface beyond traditional enterprise networks.

Surrounding the issues of security relating to the individual technology enablers there is plenty of literature which evaluates the security risks of each \cite{khan2019survey}. The biggest challenge now for industry and researchers is evaluating the holistic security risks to the 5GCN when these technologies are combined. This includes the trust boundaries both within the 5GCN and with the external services interfacing with the core network. The network architecture and vulnerabilities form the basis of traditional vulnerability based security risk assessments. Given the current maturity of the 5GCN this poses a significant issue in that traditional security risk assessments are not well suited for the current state of development and alternative approaches including threat based risk assessments may be more suitable. 
A suitable security model for 5G networks should take a holistic view of the various new and existing threats to telecommunication service providers. Particularly when considering multi-stage attack scenarios, a suitable model should consider the chain of attack steps and the consequences on the network state of each step and overall. Such a challenge requires consideration of the attackers' end goal, the threat of a network intrusion and every attack step in between. One of the most well recognized cyber security framework is that of National Institute of Standards and Technology (NIST) \cite{NISTframework}. The NIST framework core comprises five stages to manage cyber security risk which are \textit{identify, protect, detect, respond, recover}. 

With threat based risk assessments the first step of the process involves the identification of threats which helps to define the requirements for protecting the network and detection. The Process for Attack Simulation and Threat Analysis (PASTA) is a risk centric threat modelling framework \cite{ucedavelez2015risk} which consists of seven stages in which threat analysis is included at step 4. In PASTA, threat analysis includes the generation of probabilistic attack scenarios which would include the modelling of multi-stage attacks. Our first focus is on identifying suitable security monitoring within the 5GCN through threat identification and decomposition in line with the NIST framework and the threat analysis step of the PASTA framework. In the following sub-sections we apply these frameworks in the context of 5G networks. Starting with the identification of threats to the 5GCN, we review relevant literature applicable to threat identification, evaluation of probabilistic attack scenarios and the current approaches to modelling multi-stage attack scenarios.

In this chapter, we propose a framework for modelling threats for the purpose of identifying security monitoring requirements in the 5GCN which includes sensor placement and detection methods through threat decomposition and threat modelling using graphs.

The remainder of this chapter is structured as follows; in section \ref{sec:RelatedWork} we present the related work in 5GCN threat assessments and modelling of multi-stage attack scenarios. In part \ref{sec:5GThreatModelling} we propose and demonstrate a method of modelling multi-stage attack threat scenarios using graphs applicable to the 5GCN. In part \ref{sec:SecMon} we describe how threat graphs can be used to provide decision support on security monitoring through sensor placement and how this can be used for the detection of multi-stage attack threat scenarios. Finally part \ref{sec:Conclusion} of this chapter outlines the open research challenges and future work.

\section{Related Work}
\label{sec:RelatedWork}
In this section we review the literature relating to the 5GCN threats and approaches towards multi-stage attack modelling.  

\subsection{5GCN Threats}
With the introduction of new technologies to the 5GCN several 5G threat assessments have been produced \cite{khan2019survey, ahmad2018overview,rudolph2019security,ENISAThreatReport}. In these works threats are evaluated with respect to the new technologies that are introduced in 5G mobile networks including SDN, NFV and the new Service Based Architecture (SBA) within the 5G core. With a focus on the SDN and NFV technologies \cite{chen2016software} identifies the threats to NFV components such as firewalls and IDS and the interfaces between the architectural layers of the data, control and application planes introduced by SDN. One of the biggest transitions within the 5GCN is the introduction of the SBA. In this architecture the Network Functions (NF) communicate using HTTP/2 which introduces openness and flexibility but the commonality of the protocol introduces several new threats \cite{rudolph2019security}. An EU report on the threat landscape for 5G networks \cite{ENISAThreatReport} provides a taxonomy of threats and identifies which of the 5GCN infrastructure components are at risk of each threat scenario. Table \ref{table:5GThreats} provides a summary of the 5G network threats identified in the current literature.

\subsection{Multi-Stage Attack modelling}
The 5GCN threats which have been identified in literature can be considered an abstract representation of an adversary's goal realised by a multi-stage attack scenario. Initially an adversary would be required to gain access to the core network followed by several attack steps in order to achieve some goal as demonstrated by the threats in Table \ref{table:5GThreats}. Attack graphs have been researched as a solution for modelling multi-stage attacks and quantifying security risk \cite{shandilya2014use, noel2010measuring}. Attack graphs can be useful at both the design phase \cite{gupta2007using} as well as for real time intrusion detection through alert correlation \cite{ wang2006using, ahmadinejad2011hybrid, chung2013nice}. The process of automatically generating attack graphs when there is a change to the network topology or on discovery of new vulnerabilities \cite{sheyner2002automated} has become critical for many modern networks due to their dynamic nature - a feature present in the 5GCN. As attack graphs have proved valuable tools for conducting security risk analysis several tools have emerged \cite{ou2005mulval, jajodia2005topological, ingols2006practical} for automatic attack graph generation. Although drawbacks have been identified relating to readability and generation time, it is widely agreed the output proves a valuable tool for modelling attack scenarios.

\begin{table}%1
%\noautomaticrules
\caption{The key threats identified to the 5GCN}
\begin{tabular}{p{0.45\textwidth}p{0.1\textwidth}p{0.1\textwidth}p{0.12\textwidth}p{0.1\textwidth}}
{Threat Description}    &{\cite{ENISAThreatReport}} &{\cite{khan2019survey}} &{\cite{ahmad2018overview}}
&{\cite{rudolph2019security}}\\
\midrule
 Flooding attack of a NF  & \checkmark & \checkmark & \checkmark & \checkmark\\
 Eavesdropping the SBI  & \checkmark & & \checkmark & \checkmark\\
 Gaining access to data on a NF  & \checkmark & & \checkmark & \checkmark\\
 Installation of malware on NF environment  & \checkmark & & \checkmark & \\
 Modification of signaling traffic  & \checkmark & \checkmark & & \checkmark\\
 Modification of application data  & \checkmark & & & \checkmark\\
 Modification of configuration data  & \checkmark &  & & \checkmark\\
 API vulnerability exploit  & \checkmark & \checkmark & & \\
 Exploit of network protocol  & \checkmark & \checkmark & & \\
 SDN controller attack  & \checkmark & \checkmark & \checkmark & \\
 Abuse of Lawful Interception function  & \checkmark & & & \\
 Roaming scenario attacks  & \checkmark & \checkmark & \checkmark & \\
 \bottomrule
\end{tabular}
\label{table:5GThreats}
\end{table}

Attack graphs usually rely on knowledge of the network topology and known vulnerabilities within the network to identify the potential attack steps an attacker may take to achieve a specific goal. Less common than attack graphs for security risk analysis are threat graphs. Threat graphs have been used for conducting risk assessments for enterprise networks \cite{lippmann2016threat} and to evaluate the impact of threats on the security objectives of an organization to help evaluate the effect of security controls \cite{breu2008quantitative}. Although more abstract than attack graphs, threat graphs encompass a range of additional threats beyond software and hardware vulnerability exploits.

More specific to 5G networks, the authors of \cite{tian2019automated} present a graph model for multi-stage attack scenarios relating to the critical assets of the hierarchical network architecture of the 5GCN. In this work an automated attack and defence framework is proposed based on the attacker actions. Although vulnerabilities are generalised in this model rather than hardware or software specific ones it does nonetheless rely on knowledge of vulnerabilities in the 5G network. As with many other approaches, any technique which focuses on software vulnerabilities to model multi-stage attacks tend to overlook other types of threats such as insider threats, zero-day vulnerabilities or supply chain threats to name a few.

\section{Threat modelling of the 5GCN}\label{sec:5GThreatModelling}
With reference to some of the well established threat modelling frameworks such as the NIST framework for improving critical infrastructure \cite{NISTframework} and the Process for Attack Simulation and Threat Analysis (PASTA) we identify some of the key activities as part of our modelling framework and outline the hybrid approach to threat modelling for the 5GCN.

\textit{Identify}: This set of activities form part of the identification process which includes identification of: the key assets and components of the network, high level threat scenarios, network interactions such as interfaces and data flows and network boundaries. The outputs of this process include identification of the network attack surface, potential attack paths within the network and attack vectors. 

\textit{Threat Decomposition}: An important part of identifying the security monitoring requirements for threat detection is understanding the particular threat events associated with each threat scenario. This includes where within the network an attack could be launched and which components of the network may be targeted. We provide an example of the threat decomposition process in a later section.

\textit{modelling threats with graphs}: We propose the use of graphs to model threats in the 5GCN for mapping threat sources to threat targets. For the proposed model threat scenarios and 5GCN components represent graph nodes where edges between nodes represent possible threat sources and targets. We provide a further definition of the graph model in section 1.3.3.

\subsection{Identification}
For modelling threats and the associated multi-stage attacks we first require an understanding of the network topology. In traditional models this will include knowledge of specific software, services and hardware but this is approach is not suitable for the 5GCN given the maturity of development. Additionally it is anticipated that various service providers will likely use different vendors and therefore there is no single representative 5GCN deployment we can refer to. For this reason we propose using the high level network architecture according to the latest 3GPP specification for providing the baseline architecture as shown in Fig \ref{fig:Architecure}. For our 5GCN model we define the following: \\

\textbf{Components} represent a group of assets which belong to one of the sub-systems of the 5GCN providing some overarching functionality. These are the Management and Orchestration (MANO), SDN, NFV, cloud infrastructure consisting of both the physical (PI) and virtual infrastructure (VI) and the application layer virtualised Network Functions (NF) of the SBA. \\

\textbf{Threat Scenario} a set of threat events associated with one or more threat source and target. \\

\textbf{Threat Event} can be considered as a specific attack step contributing to some undesired outcome of a threat scenario.\\

\textbf{Threat Source} is the term used for the network component from where a threat event is launched. \\

\textbf{Threat Target} is the term used for the network component which is targeted by a threat event. \\

In the following sub-section we provide further details on the process of threat decomposition of threat scenarios into the threat events, source and target.

\subsection{Threat Decomposition}
The process of threat decomposition involves taking the threat scenarios identified in literature to produce a set of threat events and assigning the potential source and targets within the 5GCN. We provide a working example of the process as follows:\\

\textbf{Threat Scenario:} \textit{API vulnerability exploit}

\textbf{Threat Events:} \textit{SQL injection, buffer overflow, remote code execution}

\textbf{Threat Source:} \textit{external networks}

\textbf{Threat Target:} \textit{Network Functions, MANO} \\

In this example the threat \textit{API vulnerability exploit} as identified in the literature can be realised through the use of several different threat events, each of which is a technique an attacker may use to exploit the pubic facing API's provided to the 5GCN. Both the MANO and some of the 5GCN network functions such as the Network Exposure Function (NEF), Access Management Function (AMF) and Security Edge Protection Proxy (SEPP) provide a public facing API and therefore are potential targets of this threat scenario. These components provide an interface for external networks and systems to interact with the 5GCN and so in this example the potential threat source is the external networks. 

Beyond helping to guide decisions on security monitoring requirements for the 5GCN the process of threat decomposition also helps to identify, if detected, which stage of a multi-stage attack the threat event belongs. Although the derived threat events are not as specific as known vulnerability exploits which are used in some security risk assessments they do provide an abstract representation of attack steps. This approach shares similarities with the MITRE ATT\&CK framework which provides a set of Techniques, Tactics and Procedures (TTP) commonly used by adversaries targeting enterprise networks. The MITRE ATT\&CK framework\footnote{https://attack.mitre.org/matrices/enterprise/}defines a tactic as the goal or aim of each attack step and techniques as the method or "how" the tactic is achieved. In the process we describe the threat events can be thought of as similar to techniques i.e. how the threat scenario is realised. The tactics defined in the MITRE ATT\&CK framework represent the 12 different stages associated with multi-stage attack scenarios typical with APT campaigns. Although beyond the scope of our work, the process of assigning a specific technique or threat event to a tactic group can help in the detection of APT multi-stage attacks. We discuss related work using this approach in the concluding section of this chapter.

\subsection{Modelling 5GCN Threats as Graphs}
As a tool for modelling multi-stage attacks, attack graphs can be utilised to map the relationship between attack steps to network infrastructure. Rather than using graphs to map attack steps in respect to vulnerability exploits we investigate the use of graphs for modelling threat scenarios relating to the 5GCN infrastructure.

Threat scenarios can be mapped to the network infrastructure and graph edges between different components can help to identify where in the network security sensors should be placed in order to detect a threat event. A cyber-threat can be defined as \textit{the possibility of a malicious attempt to damage or disrupt a computer network or system}. We extend this concept to include a source and target of a threat with regards to the network components. In the context of the 5GCN this relates to the key components we have identified. Any threat scenario $T$ can be defined as a tuple $T = (src, dst, scenario)$ where $src$ is the component from where the malicious action originates and $dst$ is the destination or the component being targeted by the threat $scenario$. As a simplistic example the $T =$ (External Network, NEF, API Exploit) describes the threat of exploiting the API of the NEF from an external network. We can model the 5GCN threats in a graph form: \\

\noindent\textbf{Definition 1. Threat Graph} Given a set of threat scenarios $t \in T$, a set of 5GCN infrastructure components $c \in C$, a threat source $T_s \subseteq C \times T$ and a threat target $T_t \subseteq T \times C$ a threat graph $G$ is a directed graph $G = (C \cup T, T_s \cup T_t)$ where $C \cup T$ is the vertex set and $T_s \cup T_t$ the set of edges. \\

An important feature of the threat graph model is the relationship between multi-stage attack steps'. A component node of the graph can serve as both a threat target and if that attack step is successful provide a platform for further attacks originating from the compromised component. As a result this component will then become a threat source represented by the outbound edges in the graph. Likewise inbound edges of a network component node represent it is a threat target. This concept plays a key role in the detection of multi-stage attacks. For our threat graph model we assume the attacker starts outside of the 5GCN and as such the initial node of the graph is represented by the external network node. Graph edges starting from the external network represent the first step of a multi-stage attack and each proceeding edge from a 5GCN component to a threat scenario represents $1 + n$ attack steps.

\subsection{A 5GCN Threat Graph Example}
\begin{figure}[h]
\centering
\includegraphics[scale=0.8]{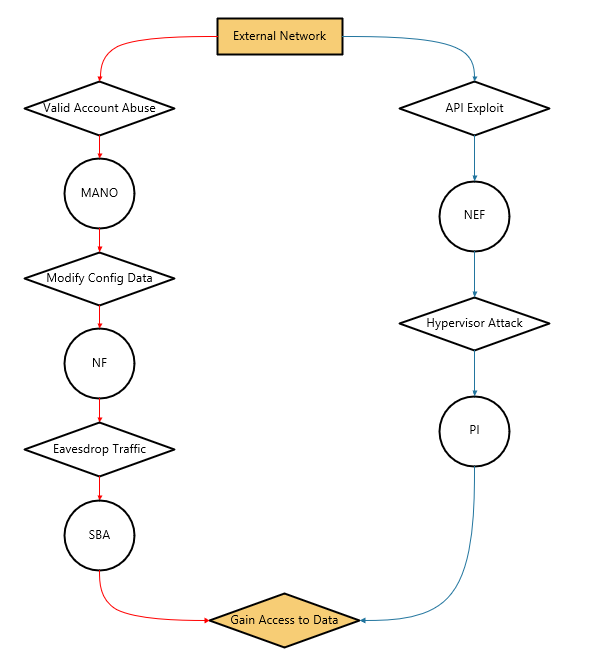}
\caption{5GCN Threat Graph}
\label{fig:ThreatGraph}
\end{figure}

With the graph representation of threats and the relationship between 5GCN infrastructure components we can represent the progression of a multi-stage attack scenario. The example threat graph in Fig \ref{fig:ThreatGraph} demonstrates how the threats identified in the literature can be related to the 5GCN components. 

From the example threat graph we identify two multi-stage attack scenarios and explain how these are represented on the graph.

\textbf{Threat Scenario 1}
\begin{enumerate}
    \item An attacker gains credentials for the MANO component
    \item The adversary is able to make configuration changes which introduces a malicious NF into the SBA.
    \item The malicious NF begins to harvest sensitive data by eavesdropping the SBA.
\end{enumerate}

\textbf{Threat Scenario 2}
\begin{enumerate}
    \item An attacker exploits a vulnerability in the API of the NEF.
    \item The attacker establishes a remote connection to the NEF of the 5GCN.
    \item The attacker exploits a vulnerability in the hypervisor and gains access to the PI.
    \item The adversary is able to access memory storing data for all of the NFs in the SBA.
\end{enumerate}

In the above scenarios the threat graph demonstrates the steps an attacker may take by moving laterally between the 5GCN components. This approach provides a holistic view of how an adversary may attack the different key components in a multi-step threat scenario without only focusing on the security events relating to the individual components in isolation. Scenario 1, highlighted by the red edges, and Scenario 2, highlighted by the blue edges, show how an adversary can attack the 5GCN through different attack vectors with the same goal of gaining access to sensitive data. With regards to security monitoring placement we can see where an attack may originate and which components could be targeted and so are provided with guidance on sensor placement to detect threat events. In section 1.4 we further discuss security monitoring requirements using the example threat graph for reference. 

\section{Graph Based Security Monitoring for the 5GCN Infrastructure}\label{sec:SecMon}
Effective security monitoring within a network includes two important aspects; first is knowing where in the network the security monitoring should be applied and second what type of security monitoring is required. With traffic volume and number of connected devices in the 5G network anticipated to surpass levels previously seen traditional approaches to security monitoring may not be suitable. Analysis of the network for applying security monitoring may quickly become outdated due to the dynamic, re-configurable nature of the core network and techniques such as deep packet inspection (DPI) of traffic for detection of threats will need to be applied only where and when required. Due to the large amounts of data transiting the 5G network many existing DPI tools would not be able to process the results in real time and may even miss suspicious packets. In this chapter we have described a process involving threat decomposition for mapping threats to 5GCN components using graphs to identify where in the network security monitoring should be placed to detect threat events. The process of threat decomposition also provides identification of threat events which gives further guidance of the methods to detect them. The last stage of the framework is to provide a set of security monitoring requirements which can be translated into a set of rules applied to the 5GCN capable of providing effective security monitoring despite the challenges posed by the dynamic nature and high data throughput. 

When considering the security monitoring of specific threats examples include the use of anti-virus software for detecting malware signatures or firewalls which can monitor traffic flows which can be used to detect malicious connections within the network. In most cases the type of security monitoring is guided by knowing what type of attack technique is used. The placement of security sensors within the network by contrast relies on knowledge of the network architecture, identification of the critical assets and the threats posed to those assets. As previously explained modelling these threats in the form of a graph provides a way of doing this for providing security monitoring requirements by identifying which threats apply to specific 5GCN components. When it comes to security monitoring and specifically Intrusion Detection Systems (IDS) there are two types that can be applied. First is Host based IDS (HIDS) which as the name suggests are responsible for the security monitoring of a specific host. With reference to the 5GCN architecture this could include the physical and virtual hosts within the network on which components like the VNF, SDN and MANO components reside. The second type of security monitoring is Network based IDS (NIDS) and is focused on monitoring traffic and connections within the network. In the context of the 5GCN and the threat graph produced traffic based security monitoring is important for the detection of threat events between two different 5GCN components. The threat graph \ref{fig:ThreatGraph} provides a mapping between threat events and the \textit{at risk} 5GCN network components for identifying \textit{where} security monitoring is required. To understand \textit{what} type of monitoring is required in these locations we further evaluate the threat events and identify network parameters to be monitored. \\

\noindent \textbf{Threat Scenario:} \textit{Flooding attack of a NF}

\noindent \textbf{Security Monitoring:} Traffic volumes, network connections, resource usage and error codes.

\noindent \textbf{Description:} There are several threat events that could contribute to a flooding attack of a NF. The detection of this threat type relies on identifying abnormal traffic flows, between the threat source and target components, which could include an increase in volume generated by user equipment such as a botnet attack or through an innocent or malicious NF generating additional traffic. In both scenarios monitoring traffic volumes and any error codes that may result from illegitimate HTTP/2 traffic can help identify this threat. \\

\noindent \textbf{Threat Scenario:} \textit{Eavesdropping the SBI}

\noindent \textbf{Security Monitoring:} n/a

\noindent \textbf{Description:} Directly detecting a NF that is eavesdropping is a difficult task. Instead it is advised to monitor the symptoms of eavesdropping such as data collection of intercepted communication through abnormal file sizes or types. Additionally monitoring registered NFs for suspicious behavior can help identify malicious NFs registered in the SBA. \\

\noindent \textbf{Threat Scenario:} \textit{Gaining access to data on a NF}

\noindent \textbf{Security Monitoring:} Monitor access to data repositories 

\noindent \textbf{Description:} A lot of sensitive data in the 5GCN is stored in data repositories. Detection of an adversary accessing the data will rely on monitoring access particularly attempted access which is denied by the system. Properties such as frequency of access and size of data requests should be monitored for abnormalities. \\

\noindent \textbf{Threat Scenario:} \textit{Installation of malware on NF environment}

\noindent \textbf{Security Monitoring:} anti-virus software, monitor system logs of new software and monitor behaviour such as resource usage and system calls for anomalies.

\noindent \textbf{Description:} In most cases the use of anti-virus software is deemed adequate in the detection of malware. When considering APTs and the efforts taken to remain undetected, the use of any malware is likely to take additional efforts to evade defences. An adversary may disguise malware in an applications or executable, deploy zero day attacks or even introduce a malicious NF to the 5GCN which could bypass anti-virus software. Therefore in addition to anti-virus software it may be necessary to monitor the installation of new applications or software and the behaviour of them to identify any malicious side effects. \\

\noindent \textbf{Threat Scenario:} \textit{ Modification of application data}

\noindent \textbf{Security Monitoring:} Monitor system logs 

\noindent \textbf{Description:} Modification of application data may be an intended action but it could also indicate malicious activity. Although not definitively a signal of malicious activity system logs should be used to raise alerts when application data is modified including information such as the user that made changes. \\

\noindent \textbf{Threat Scenario:} \textit{Modification of configuration data}

\noindent \textbf{Security Monitoring:} Monitor system logs

\noindent \textbf{Description:} Modification of configuration data within the 5GCN is anticipated to be normal behaviour. Routing tables, network slicing and virtualised network functions will be re-configurable and dynamic in nature to support the network demands. Adversaries may take advantage of this feature for redirection of traffic or introduce vulnerabilities to the configuration of NFs. \\

\noindent \textbf{Threat Scenario:} \textit{API vulnerability exploit}

\noindent \textbf{Security Monitoring:} Monitor application logs

\noindent \textbf{Description:}  Monitoring logs of the public facing applications providing APIs for behavioural abnormalities may help to detect attempted or successful exploits. With large volumes of traffic anticipated within the 5GCN, DPI may not be effective for real-time analysis. The use of application firewalls can be deployed to detect invalid inputs as an indicator of attempted exploits. \\

\noindent \textbf{Threat Scenario:} \textit{SDN controller attack}

\noindent \textbf{Security Monitoring:} monitor system logs, resources usage and access, traffic flows

\noindent \textbf{Description:} The SDN controller plays a critical role in the 5GCN. Attacks on the SDN controller range from DoS type attacks, malicious configuration changes to redirect traffic and for gaining knowledge of routing tables. A range of monitoring should be used to detect changes but also to monitor symptoms of any malicious activity. \\

\noindent \textbf{Threat Scenario:} \textit{Abuse of Lawful Interception (LI) function}

\noindent \textbf{Security Monitoring:} monitor use of LI function

\noindent \textbf{Description:} Monitoring of abuse of the LI function is not an easy task. Monitoring when the LI function is used and verifying it with the relevant authorities is the most reliable method of detection. With regards to the LI function being used via a visited mobile network operator maintaining a whitelist or blacklist of trusted/untrusted visited networks may help to identify misuse. \\

\noindent \textbf{Threat Scenario:} \textit{Roaming scenario attacks}

\noindent \textbf{Security Monitoring:} Blacklisted networks, correlate user location tracking

\noindent \textbf{Description:} There are several threat events posed by the roaming scenario threat. There are several methods which could be used to detect roaming scenario abuse including maintaining a blacklist of fraudulent visited network requests and geographical tracking of user equipment to check if the location matches expected location of a user. \\

The detection methods identified provide ways of generating alerts of potential suspicious behaviour. In many cases these alerts could be innocent and normal operation but the more threat events that are detected and then correlated the probability of detecting a multi-stage attack increases whilst the false positive rate of detection should be reduced. Many alerts such as suspicious user logon activity or reconfiguration of the network are expected to be 
We discuss the process of alert filtering and correlation in the future work section of this chapter.

\section{Open Research Challenges and Future Work}\label{sec:Conclusion}
In this chapter we describe a framework for the detection of multi-stage attacks on 5G networks and in particular the 5GCN. The framework takes a three stage approach to translate high level threat scenarios and model these as threat events against the 5GCN infrastructure components using graphs. In this section we discuss the challenges and future work for further developing the framework for the detection of APTs in 5G networks.

\subsection{Detection of Multi-Stage Attack Scenarios}
Different approaches have been evaluated for detecting multi-stage attacks based on security alerts. The merits of correlating security alerts to the "attack lifecycle" involving the multiple tactic groups are presented in \cite{ghafir2018detection}. The idea is that during a multi-stage attack the different tactics would be deployed in chronological order. Starting with a network intrusion and usually resulting in some data being exfiltrated from the network. An attacker would deploy tactics, as identified in the MITRE Att\&ck framework, to remain undetected whilst moving laterally around the network, gaining the required privileges to gain access and gather data before finally transferring it out of the network, sometimes by establishing Control and Command (C\&C) node to hide the process. Through this chronology of events it is possible to correlate security alerts by the order in which they occur and to which tactic group they belong. For a security event that could indicate a suspicious user login followed by some event relating to a malicious configuration change within the network by the same user account there is a potential for these two events to be related. This concept, as with many attack graph models, assumes a prerequisite action or state to occur beforehand. In the case of the threat graph this same assumption exists. If we consider the threat of a signalling attack in the Service Based Interface (SBI) where one NF is a source and another is the target we assume the source component has been compromised and under the control of the attacker.

In addition to relating security events based on the order in which they occur and to which tactic group they belong the threat source and target can be used to further correlate security events. When a security alert is generated the source, destination and alerts type can be translated into the nodes and connecting edges of the threat graph. If we consider an alert $A = (src, dest, alert)$ this can be translated into the connection from a threat source $(component, threat)$ to the threat target $(threat, component')$ where $component = src$, $threat = alert$ and $component' = dest$ and matches can be found through a graph traversal.

Through this process of security alert correlation we propose the following steps for the detection of multi-stage attacks:

\begin{itemize}
    \item \textbf{Chronology of Events} this step involves sorting the security alerts by the order in which they occur.
    \item \textbf{Identification of Attack Tactic} once security alerts have been ordered each event is related to a specific attack stage aligned to the tactic group they belong.
    \item \textbf{Relating Events to 5GCN Infrastructure} finally the filtered security alerts are correlated based on the network infrastructure components which are the source or target. This stage of correlation looks to identify components which are firstly targeted by an attack which later become the source of an attack step.
\end{itemize}

The three fold approach to security alert filtering and correlation may provide a solution to issues relating to overlooked security alerts. Security alerts analysed in isolation may not provide a holistic view of the multi-stage attack steps, however, when analysed in the context of attack tactics and the infrastructure components affected the process of linking different events becomes a useful tool.

\subsection{A Centralised Threat Detection Engine}
The process of decomposing threats, linking the threats to the 5GCN infrastructure components and identifying suitable security monitoring requirements and for the placement and detection of threat events is a first step and can be conducted without detailed network data. To evaluate the feasibility and performance of detecting threats in a centralised manor requires the various vendors of MANO, SDN, NFV and cloud computing to provide the security monitoring data suggested in a common way. There are multiple vendors who provide solutions such as virtualization software suitable for deployment in 5G networks each of which may have different security monitoring capabilities for their products. Access to any security monitoring data for the purpose of threat detection requires this to be made available for analysis of a centralised threat detection engine.

The future development based on the framework described will rely on collating the security monitoring data including system logs and traffic flows being made available by vendors. In order to test the process described for analysing security alerts a quality set of data will be required consisting of both genuine and false positive alerts to measure the effectiveness of the process. 

\subsection{A 5G Network TTP Matrix}
In the framework proposed we describe how threat scenarios can be decomposed into threat events which represent the techniques that the security monitoring system should be capable of detecting. The process of detecting threat events will rely heavily on a full and comprehensive list of adversarial techniques used to attack 5G networks. The MITRE ATT\&CK framework for enterprise networks is the result of threat intelligence based on real-world observations. It may be possible to provide a similar TTP framework for 5G networks but this would require a similar level of cooperation between mobile networks to disclose information about network attacks which may prove challenging. The threats which have been identified in literature relating to the new 5G technologies provide a platform for collating the different techniques applicable to 5G networks which could prove valuable future work for the initial development of such a framework but remains an open research challenge.

\bibliographystyle{unsrtnat}
\bibliography{references}  %%% Uncomment this line and comment out the ``thebibliography'' section below to use the external .bib file (using bibtex) .





% \end{thebibliography}

\end{document}